\documentclass[12pt]{article}
\usepackage{amsmath}
\begin{document}
\begin{flushright}
KANAZAWA-15-13\\
July, 2015
\end{flushright}
\vspace*{1cm}

\begin{center}
{\Large\bf Lepton number asymmetry via inflaton decay\\ 
in a modified radiative seesaw model}
\vspace*{1cm}

{\Large Shoichi Kashiwase}\footnote[1]{e-mail:~shoichi@hep.s.kanazawa-u.ac.jp} 
{\Large and}
{\Large Daijiro Suematsu}\footnote[2]{e-mail:~suematsu@hep.s.kanazawa-u.ac.jp}
\vspace*{1cm}\\

{\it Institute for Theoretical Physics, Kanazawa University, \\ 
    Kanazawa 920-1192, Japan}
\end{center}
\vspace*{1.5cm} 

\noindent
{\Large\bf Abstract}\\
We propose a non-thermal scenario for the generation of 
baryon number asymmetry in a radiative neutrino mass model 
which is modified to realize inflation at the early Universe.
In this scenario, inflaton plays a crucial role in
both generation of neutrino masses and lepton number asymmetry.
Lepton number asymmetry is firstly generated in the 
dark matter sector through direct decay of inflaton.
It is transferred to the lepton sector via the dark matter 
annihilation and then converted to the baryon number asymmetry 
due to the sphaleron interaction.  
All of the neutrino masses, the baryon number asymmetry and 
the dark matter are intimately connected to each other 
through the inflaton.
\newpage
\section{Introduction}
Recent experimental and observational data for neutrino masses \cite{nexp,t13}
and dark matter (DM) \cite{uobs,wmap,planck} suggest that 
the standard model (SM) should be extended.
The radiative neutrino mass model proposed in \cite{ma} is such a simple 
extension of the SM with an inert doublet scalar and right-handed neutrinos.
It seems to be a promising candidate which could take the place of 
the famous canonical seesaw model for neutrino masses \cite{seesaw}.  
An interesting point of this model is that it could also give the origin 
of DM \cite{raddm1,raddm2}.
A $Z_2$ symmetry imposed to forbid the neutrino masses at tree-level
could guarantee the stability of the lightest $Z_2$ odd field,
which could be DM. In this model, DM is an indispensable ingredient 
for the neutrino mass generation at TeV regions.

Although the model has such interesting aspects, baryon number asymmetry 
in the Universe \cite{bbn}, which is another crucial problem of the SM, 
cannot be easily explained in a consistent way with the relic
abundance of DM. 
If we suppose the ordinary thermal leptogenesis \cite{fy,leptg}, 
the sufficient baryon number asymmetry can be generated only in the case
where the model has a finely tuned spectrum for the $Z_2$ odd fields.

If the lightest right-handed neutrino is assumed to be DM,
both its relic abundance and small neutrino masses require 
$O(1)$ neutrino Yukawa couplings in general\footnote{
This brings about dangerous lepton number violating processes at
large rate unless special flavor structure is assumed for 
the neutrino Yukawa couplings \cite{raddm2}.} \cite{raddm1}. 
They can allow to cause large $CP$ asymmetry in the decay of
right-handed neutrinos even if their masses are of $O(1)$~TeV.
However, the same neutrino Yukawa couplings could cause large washout 
of the generated lepton number asymmetry through the inverse decay and 
the lepton number violating scattering processes.
As a result, the thermal leptogenesis is not easy to generate sufficient 
lepton number asymmetry in a consistent way with the neutrino
oscillation data and the DM abundance at least in the simplest form 
of the model \cite{radlept}. 
On the other hand, if the lightest neutral component of the inert 
doublet scalar is assumed to be DM \cite{ham}, the neutrino 
Yukawa couplings could 
be small enough to be consistent with both the DM relic abundance 
and the small neutrino masses. However, the large $CP$ asymmetry in the 
decay of right-handed neutrinos requires fine mass degeneracy 
among the right-handed neutrinos \cite{ks}. 
Non-thermal leptogenesis \cite{ad, inf-nradm} might give another 
consistent scenario for the origin of the baryon number asymmetry 
in this model or its supersymmetric extension \cite{susyrad}. 

In this paper, to solve the above mentioned
fault for leptogenesis, we propose a simple scenario in the model 
which is extended so as to incorporate the inflation 
at the early Universe \cite{bks}. 
The neutrino mass generation is connected with the inflation 
through the inflaton interaction. 
The lepton number asymmetry is also produced through the 
inflaton decay in the inert doublet sector which contains 
the DM candidate \cite{inf-nradm,bks}.  
After this lepton number asymmetry is transferred to the lepton sector 
via lepton number conserving scattering processes, 
the sphaleron interaction converts a part of it to the 
baryon number asymmetry.

Remaining parts of the paper are organized as follows. 
In the next section, we introduce the 
extended model briefly. In section 3, we study its phenomenological 
features. Firstly, we describe the inflation in the model and 
also the small neutrino mass generation. After that, we explain the scenario
for the generation of the lepton number asymmetry and then estimate 
the baryon number asymmetry expected to be produced finally.  
Following this discussion, the consistency of the scenario 
with DM phenomenology is examined.
Relation between the present DM scenario and the asymmetric DM scenario 
is also remarked. We summarize the paper in section 4.

\section{An extension of the radiative seesaw model}
Our model considered here is based on the one proposed for the 
radiative neutrino mass generation \cite{ma}.
The original model is a simple extension of the SM 
with an inert doublet scalar $\eta$ and three right-handed neutrinos 
$N_{R_i}$. These new fields are assigned odd parity of an 
imposed $Z_2$ symmetry, although all the SM contents are assumed to have its 
even parity. Invariant Yukawa couplings and scalar potential which are 
relevant to these new fields are summarized as
\begin{eqnarray}
-{\cal L}_y&=&h_{ij} \bar N_{R_j}\eta^\dagger\ell_{L_i}
+h_{ij}^\ast\bar\ell_{L_i}\eta N_{R_j}
+\frac{1}{2}\left(M_i\bar N_{R_i}N_{R_i}^c 
+M_i\bar N_{R_i}^cN_{R_i}\right), \nonumber \\
&+&m_\phi^2\phi^\dagger\phi+m_\eta^2\eta^\dagger\eta
+\lambda_1(\phi^\dagger\phi)^2+\lambda_2(\eta^\dagger\eta)^2
+\lambda_3(\phi^\dagger\phi)(\eta^\dagger\eta) 
+\lambda_4(\eta^\dagger\phi)(\phi^\dagger\eta) \nonumber \\
&+&\frac{\lambda_5}{2}\left[(\phi^\dagger\eta)^2 +{\rm h.c.}\right],
\label{model}
\end{eqnarray}
where $\ell_{L_i}$ is a left-handed doublet lepton and $\phi$ is 
an ordinary doublet Higgs scalar. We use the basis for which both matrices 
for charged lepton Yukawa couplings and right-handed neutrino masses 
are real and diagonal.
Since the $Z_2$ is assumed to be the exact symmetry of the model, 
the new doublet scalar $\eta$ should not have a vacuum expectation value.
As its result, neutrino masses are forbidden at tree level
and the lightest field with the odd parity is stable to be DM. 

In this type of model, the lepton number $L$ is usually assigned to 
these new fields as $ L(\eta)=0$ and $L(N_{R_i})=1$.
In such a case, the neutrino mass generation and leptogenesis have been 
studied under the assumption that mass terms of the right-handed neutrinos 
violate the lepton number \cite{radlept,ks}. 
The DM abundance has also been studied supposing that 
either the lightest right-handed neutrino or the lightest 
neutral component of $\eta$ is DM. 
However, it is useful to note that there could be another assignment 
of the lepton number such as $ L(\eta)=1$ and $L(N_{R_i})=0$ \cite{inf-nradm}. 
In this case, $\lambda_5(\phi^\dagger\eta)^2$ is forbidden 
as long as the lepton number is imposed as the exact symmetry.
As a result, neutrino masses
could not be generated even if the radiative effect is taken into account.
Thus, some suitable origin of the lepton number violation should bring about 
this $\lambda_5$ term as an effective interaction at low energy regions. 
We study such a possibility in the following part.

For this purpose, we consider an extension of the model at high energy regions
by introducing canonically normalized complex singlet scalars $S_\alpha$ 
which are assigned odd parity of the $Z_2$ symmetry and $L=1$. 
The potential and interaction terms of $S_\alpha$ are assumed 
to be given by 
\begin{eqnarray}
-{\cal L}_S&=& \sum_{\alpha=1}^2\Big(
\kappa_1(S_\alpha^\dagger S_\alpha)^2 
+\kappa_2(S_\alpha^\dagger S_\alpha)(\phi^\dagger\phi)
+ \kappa_3(S_\alpha^\dagger S_\alpha)(\eta^\dagger\eta)
\nonumber \\
&+& \tilde m_{S_\alpha}^2S_\alpha^\dagger S_\alpha 
+ \frac{1}{2}m_{S_\alpha}^2S_\alpha^2 +\frac{1}{2} m_{S_\alpha}^2S_\alpha^{\dagger 2}
-\mu_\alpha S_\alpha\eta^\dagger\phi - \mu_\alpha^\ast S_\alpha^\dagger 
\phi^\dagger\eta \Big)
\nonumber \\ 
&+& c_1\frac{(S_1^\dagger S_1)^n}{M_{\rm pl}^{2n-4}}
\left[1+ c_2\left\{ \left(\frac{S_1}{M_{\rm pl}}\right)^{2m} 
\exp\left(i\frac{S_1^\dagger S_1}{\Lambda^2}\right)
+ \left(\frac{S_1^\dagger}{M_{\rm pl}}\right)^{2m}
\exp\left(-i\frac{S_1^\dagger S_1}{\Lambda^2}\right)
\right\} \right], \nonumber \\
\label{model1}
\end{eqnarray}
where both $n$ and $m$ in the third line are positive integers 
and $M_{\rm pl}$ is the reduced Planck mass.
Although the $Z_2$ is kept as the symmetry of these terms, the lepton number
is violated through the mass terms $m_{S_\alpha}^2S_\alpha^2$,
$m_{S_\alpha}^2S_\alpha^{\dagger 2}$ in the second line and also the 
Planck suppressed $c_2$ terms in the third line.
The latter one is neglected in the low energy region.  
On the other hand, the former lepton number violation could be 
an origin of $\lambda_5$ term in eq.~(\ref{model}). 
In fact, as a simplest case, we might consider the situation 
where $\tilde m_{S_\alpha}^2\gg m_{S_\alpha}^2$ is satisfied. 
In this case, the model defined by eq.~(\ref{model}) can be 
easily obtained as the effective one with 
$\lambda_5=\sum_\alpha \lambda_5^{(\alpha)}$, 
where $\lambda_5^{(\alpha)}$ is defined by 
$\lambda_5^{(\alpha)}= \frac{m_{S_\alpha}^2\mu_\alpha^2}{\tilde m_{S_\alpha}^4}$. 
They are induced as the effective interaction terms at low energy regions 
after the singlet scalars $S_\alpha$ are integrated out \cite{inf-nradm,bks}.

In the following discussion, we are focus our study on the situation 
such that the terms in the last line in eq.~(\ref{model1}) could be 
a dominant part of the potential at the early Universe.
We suppose that $|S_1|$ takes a large but sub-Planckian 
value in such a period. It could be realized under 
the condition such as\footnote{When $S_1$ plays a role of inflaton, 
this condition could be relevant to the $\eta$ problem in 
this inflation scenario. We cannot fix it at this stage unless 
the UV completion of the model is clarified.}  
\begin{equation}
\kappa_1\ll c_1\left(\frac{\varphi_1}{M_{\rm pl}}\right)^{2n-4}, \qquad
\left(\frac{\tilde m_{S_1}}{\varphi_1}\right)^2,~
\left(\frac{m_{S_1}}{\varphi_1}\right)^2 \ll 
c_1\left(\frac{\varphi_1}{M_{\rm pl}}\right)^{2n-4},
\label{cond}
\end{equation}
where $\varphi_1$ is defined by $S_1=\frac{\varphi_1}{\sqrt 2}e^{i\theta_1}$ 
and $\varphi_1<M_{\rm pl}$.
If we use the polar coordinate of $S_1$ defined here,
the last line of eq.~(\ref{model1}) can be written as
\begin{equation}
V_{S_1}=c_1\frac{\varphi_1^{2n}}{2^nM_{\rm pl}^{2n-4}}\left[
1+ 2c_2\left(\frac{\varphi_1}{\sqrt 2 M_{\rm pl}}\right)^{2m}
\cos\left(\frac{\varphi_1^2}{2\Lambda^2}+2m\theta_1\right)\right].  
\label{infpot}
\end{equation}
We easily find that $V_{S_1}$ has local minima with the potential barrier
$V_b\simeq \frac{c_1c_2\varphi_1^{2(n+m)}}{2^{n+m-2}M_{\rm pl}^{2(n+m-2)}}$
in the radial direction, which form a spiral-like trajectory. 
We consider the inflation which is caused by the inflaton evolution 
along this trajectory.

\section{Phenomenological features of the model}
\subsection{Inflation}
We briefly review the features of the inflation induced by the
potential (\ref{infpot}).  
We assume that $\varphi_1$ takes a large initial value 
on a local minimum in the radial direction. 
In that case, as shown in \cite{bks},
the model could cause sufficient $e$-foldings through 
the inflaton evolution along the spiral-like trajectory 
even for sub-Planckian values of $\varphi_1$. 
An inflaton field $\chi$ could be identified with
\begin{equation}
\chi\equiv a_e+\frac{\varphi_{1e}^3}{6m\Lambda^2}-a=
\frac{\varphi_1^3}{6m\Lambda^2},
\end{equation}
where the field $a$ is defined as
\begin{equation}
da=\left[\varphi_1^2+\left(\frac{d\varphi_1}{d\theta_1}
\right)^2\right]^{1/2}d\theta
=\left[1 + 4m^2\left(\frac{\Lambda}{\varphi_1}\right)^4\right]^{1/2}
\varphi_1 d\theta_1.
\label{infl}
\end{equation}
Fields with the subscript $e$ stand for the fields at the end of inflation. 
The number of $e$-foldings caused by $\chi$ is given as
\begin{equation}
N=-\frac{1}{M_{\rm pl}^2}\int_\chi^{\chi_e} d\chi ~\frac{V_{S_1}}{V_{S_1}^\prime}
\equiv N(\chi)-N(\chi_e), 
\label{efold0}
\end{equation}
where $V_{S_1}^\prime=\frac{dV_{S_1}}{d\chi}$ and 
$N(\chi)$ is represented by using the
hypergeometric function $F$ as
\begin{eqnarray}
N(\chi)&=&\frac{1}{6m^2n}
\left(\frac{M_{\rm pl}}{\Lambda}\right)^4
\left(\frac{\varphi_1}{\sqrt 2M_{\rm pl}}\right)^6\left[~1 
+\frac{6c_2m}{n(3+m)}\left(\frac{\varphi_1}{\sqrt 2M_{\rm pl}}\right)^{2m}
\right. \nonumber \\
&&\hspace*{2cm}\left. 
\times F\left(1,~\frac{3}{m}+1,~\frac{3}{m}+2,~2c_2\left(1+\frac{m}{n}\right)
\left(\frac{\varphi_1}{\sqrt 2M_{\rm pl}}\right)^{2m}\right)\right].
\label{efold}
\end{eqnarray} 

Here we note that the model could have a different feature from 
the ordinary inflation scenario such as the chaotic inflation.
In eq.~(\ref{efold0}), $N(\chi)\gg N(\chi_e)$ might not be satisfied generally.
In this model, inflation is expected 
to end at the time when $\frac{1}{2}\dot{\chi}^2\simeq V_b$ is satisfied.
If we apply the slow-roll approximation $3H\dot\chi=-V_{S_1}^\prime$
to the one of slow-roll parameters $\varepsilon\equiv\frac{M_{\rm pl}^2}{2}
\left(\frac{V_{S_1}^\prime}{V_{S_1}}\right)^2 $ \cite{slowroll},
the inflation is found to end at $\varepsilon=\frac{3V_b}{V_{S_1}}$.
This means that the end of inflation could happen much before 
the time when $\varepsilon\simeq 1$ is realized since $V_{S_1}>V_b$ is satisfied.
In that case, $N(\chi_e)$ could have a substantial contribution 
to determine the $e$-foldings $N$ in eq.~(\ref{efold0}).  

\begin{figure}
\begin{center}
\begin{tabular}{cccccccc}\hline
$c_1$ & $c_2$ & $\frac{\Lambda}{M_{\rm pl}}$ &
$\frac{\varphi_1^\ast}{\sqrt 2 M_{\rm pl}}$ & $H_\ast$&
 $N_\ast$ & $n_s$ & $r$ \\ 
$(\times 10^{-7})$ &&&& $(\times 10^{14}{\rm GeV})$ && \\ \hline
9.84&1.7 &0.05& 0.411& 5.91& 60.0 & 0.964&0.056 \\
8.62&1.9&0.05 & 0.406& 5.40& 60.0& 0.959&0.040\\ \hline
\end{tabular}
\end{center}
\vspace*{-1mm}

{\footnotesize {\bf Table~1.} Examples of the predicted values for the 
spectral index $n_s$ and the tensor-to-scalar ratio $r$ 
in this scenario fixed by $n=3$ and $m=1$.  }
\end{figure}
 
The slow-roll parameters $\varepsilon$ and 
$\eta\equiv M_{\rm pl}^2\left(\frac{V_{S_1}^{\prime\prime}}{V_{S_1}}\right)$ 
can be represented by using the model parameters as
\begin{eqnarray}
&&\varepsilon=m^2\left(\frac{\sqrt 2M_{\rm pl}}{\varphi_1}\right)^6
\left(\frac{\Lambda}{M_{\rm pl}}\right)^4\left[
\frac{n-2c_2(m+n)\left(\frac{\varphi_1}{\sqrt 2M_{\rm pl}}\right)^{2m}}
{1-2c_2\left(\frac{\varphi_1}{\sqrt 2M_{\rm pl}}\right)^{2m}}\right]^2, \nonumber\\
&&\eta=m^2\left(\frac{\sqrt 2M_{\rm pl}}{\varphi_1}\right)^6
\left(\frac{\Lambda}{M_{\rm pl}}\right)^4
\frac{n(2n-3)-2c_2(m+n)(2m+2n-3)\left(\frac{\varphi_1}{\sqrt 2M_{\rm pl}}
\right)^{2m}}
{1-2c_2\left(\frac{\varphi_1}{\sqrt 2M_{\rm pl}}\right)^{2m}}. \nonumber \\  
\label{slow}
\end{eqnarray}
If $c_2$ terms are neglected in these formulas, we find very 
simple formulas for these slow-roll parameters at the period 
characterized by the inflaton value $\chi_\ast$.
They can be represented by using the $e$-foldings $N_\ast$
defined for $N(\chi_\ast)$ in eq.~(\ref{efold}) as 
\begin{equation}
\varepsilon\simeq\frac{n}{6(N_\ast+N(\chi_e))}, \qquad 
\eta\simeq\frac{2n-3}{6(N_\ast+N(\chi_e))}.
\end{equation}
Thus, the scalar spectral index $n_s$ and the tensor-to-scalar ratio
$r$ can be derived as \cite{bks}
\begin{equation}
n_s=1-6\epsilon+2\eta\simeq 1-\frac{n+3}{3(N_\ast+N(\chi_e))}, \qquad 
r=16\epsilon\simeq \frac{8n}{3(N_\ast+N(\chi_e))}.
\end{equation}

If we focus on the case $n=3$, these formulas reduce to the ones of the 
$m_\varphi^2\varphi^2$ chaotic inflation scenario \cite{mac}.    
However, as shown in \cite{bks}, the values of $n_s$ and $r$ in this
model could deviate from the ones of the $m_\varphi^2\varphi^2$ chaotic 
inflation due to the non-negligible $c_2$ term contribution. 
Taking account of uncertainty caused by the reheating process and others,
$N_\ast$ might be considered to take a value in the range 50 - 60. 
If we estimate both $n_s$ and $r$ by fixing the 
parameters in the potential suitably, they could take consistent 
values for $N_\ast$ in this range with the ones suggested by a joint 
analysis of BICEP2, Keck Array and Planck \cite{bkp,planck15}.
Such examples for $n=3$ are shown in Table~1. The condition (\ref{cond})
requires $\tilde m_{S_1}\ll 10^{14}$~GeV in this case.  
Much better agreement with the observational results for $n_s$ and $r$ 
is found in the case $n=1,2$ \cite{bks}. 

Finally, we note that the polar coordinate cannot be used for $S_1$ 
to rewrite the potential as eq.~(\ref{infpot}) 
unless $m_{S_1}^2=0$ is satisfied. 
In order to make this inflation scenario possible, 
$m_{S_1}^2$ should be generated after the end of inflation at least.
It is not difficult to modify the model to satisfy this condition.
For example, we may introduce a singlet scalar $\psi$ with $L=-1$.
In this case, its potential might be given by 
\begin{equation}
V_\psi=\xi_1(\psi^\dagger\psi)^2+(\xi_2S_1^\dagger S_1-m_\psi^2)
\psi^\dagger\psi 
+(\xi_3S_\alpha^2\psi^2+ {\rm h.c.}).
\end{equation}
If the value of $|S_1|$ becomes smaller than $\sqrt{\frac{m_\psi^2}{\xi_2}}$
after the end of slow-roll inflation, $\psi$ could get the vacuum 
expectation value which induces the required mass term for 
$S_\alpha$ through the $\xi_3$ term. 
After the generation of these terms in eq.~(\ref{model1})
as the effective ones, the mass splitting between 
the real and imaginary components of $S_\alpha$ is brought about. 
Each mass eigenvalue is expressed as 
$m_{\pm\alpha}^2\equiv\tilde m_{S_\alpha}^2 \pm m_{S_\alpha}^2$, 
where $+$ and $-$ signs correspond to the real and imaginary component, 
respectively. We note that the stability of the vacuum requires 
$\tilde m_{S_\alpha}^2>m_{S_\alpha}^2$.
The difference of these mass eigenvalues can 
be a measure of the lepton number violation in the model. 

\subsection{Neutrino masses}
The neutrino masses are generated in the similar way to 
the original model.
The one-loop effect which picks up the lepton number 
violation induced by the mass term $m_{S_\alpha}^2S_\alpha^2$ generates 
the neutrino masses through the electroweak symmetry breaking 
as shown in the left-hand diagram of Fig.~1.
The neutrino mass matrix obtained in this way can be described 
by the formula 
\begin{equation}
({\cal M}_\nu)_{st}=\sum_{k=1}^3\sum_{\alpha=1,2}\sum_{f=\pm}
\frac{h_{s k}h_{t k}M_k\mu_\alpha^{(f)2}\langle\phi\rangle^2}{8\pi^2}
I(M_{\eta}, M_k, m_{f\alpha}), 
\label{nmtr2}
\end{equation}
where $M_\eta^2=m_\eta^2+(\lambda_3+\lambda_4)\langle\phi\rangle^2$ and
$\langle\phi\rangle=174$~GeV. $\mu_\alpha^{(f)}$ stands for 
$\mu_\alpha^{(+)}=\frac{\mu_\alpha}{\sqrt 2}$ 
and $\mu_\alpha^{(-)}=\frac{i\mu_\alpha}{\sqrt 2}$, respectively.
The function $I(m_a,m_b,m_c)$ is defined as
\begin{eqnarray}
I(m_a,m_b,m_c)&=&\frac{(m_a^4-m_b^2m_c^2)~\ln m_a^2}
{(m_b^2-m_a^2)^2(m_c^2-m_a^2)^2}+
\frac{m_b^2~\ln m_b^2}
{(m_c^2-m_b^2)(m_a^2-m_b^2)^2} \nonumber\\
&+&\frac{m_c^2~\ln m_c^2}
{(m_b^2-m_c^2)(m_a^2-m_c^2)^2}-
\frac{1}{(m_b^2-m_a^2)(m_c^2-m_a^2)}.
 \label{mnu2}
\end{eqnarray} 
As long as $m_{\pm\alpha}^2, M_k^2 \gg M_\eta^2$ is satisfied, 
this formula is found to be reduced to
\begin{equation}
{\cal M}^\nu_{st}\simeq \sum_{k=1}^3
\frac{h_{sk}h_{tk}\langle\phi\rangle^2}
{16\pi^2M_k}
\sum_{\alpha=1,2}\left(\frac{\mu_\alpha^2}{m_{+\alpha}^2}-
\frac{\mu_\alpha^2}{m_{-\alpha}^2}\right), 
\end{equation}
where we neglect logarithmic factors.
If we note that two right-handed neutrinos are enough to explain the 
neutrino oscillation data, $h_1$ could be assumed to be so small 
that the contribution of $N_1$ to the neutrino masses is negligible. 
We adopt this assumption throughout the following discussion, for simplicity.

If we assume the flavor structure of the neutrino Yukawa couplings 
discussed in Appendix A, the required mass difference for the atmospheric 
neutrinos and the solar neutrinos could be explained by the 
largest mass eigenvalue and the next one in this mass matrix, 
respectively\footnote{It should be noted that one of the eigenvalues 
of this assumed mass matrix is zero. It may be also useful to recall that 
the cosmological upper bound for the neutrino masses 
is 0.23~eV \cite{planck15}.}.
For example, this requirement could be represented as
\begin{eqnarray}  
&&\sum_{\alpha=1,2}\left(\frac{\mu_\alpha^2}{m_{-\alpha}^2}-
\frac{\mu_\alpha^2}{m_{+\alpha}^2}\right)\simeq 
10^{-6}\left(\frac{5.1\times 10^{-2}}{h_2}\right)^2
\left(\frac{M_2}{2\times 10^4{\rm GeV}}\right),
\nonumber \\
&&\sum_{\alpha=1,2}\left(\frac{\mu_\alpha^2}{m_{-\alpha}^2}
-\frac{\mu_\alpha^2}{m_{+\alpha}^2}\right)\simeq 
10^{-6}\left(\frac{2.7\times 10^{-2}}{h_3}\right)^2
\left(\frac{M_3}{5\times 10^4{\rm GeV}}\right),
\label{nmass}
\end{eqnarray}
where we assume $M_\eta=1$~TeV and $CP$ phases are 
neglected in this estimation. 
It should be noted that the left-hand side of eq.~(\ref{nmass}) 
corresponds to the effective coupling $\lambda_5$. 
It plays a crucial role also in the generation of baryon number asymmetry
and DM direct search as discussed later.  

\input epsf
\begin{figure}[t]
\begin{center}
\epsfxsize=14cm
\leavevmode
\epsfbox{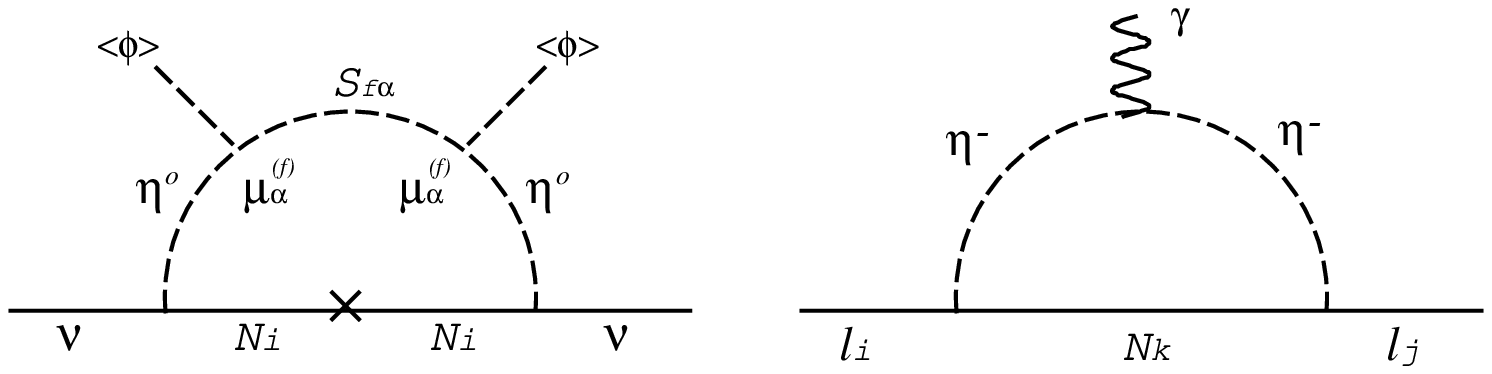}
\end{center}
\vspace*{-3mm}

{\footnotesize {\bf Fig.~1}~~Left: a one-loop diagram contributing to the 
neutrino mass generation. The dimensionful coupling $\mu_\alpha^{(\pm)}$ 
is defined as $\mu_\alpha^{(+)}=\frac{\mu_\alpha}{\sqrt 2}$ 
and $\mu_\alpha^{(-)}=\frac{i\mu_\alpha}{\sqrt 2}$ by 
using $\mu_\alpha$ in eq.~(\ref{model1}).  Right: a one-loop diagram 
contributing to the lepton flavor violating process 
$\ell_i\rightarrow \ell_j\gamma$. }
\end{figure}

It is well-known that these new fields induce the lepton flavor violating 
processes at one-loop level. The typical one is 
$\ell_i\rightarrow \ell_j\gamma$ 
whose diagram is shown in the right-hand side of Fig.~1.
Its branching ratio can be estimated as \cite{bmeg}
\begin{eqnarray}
Br(\ell_i\rightarrow \ell_j\gamma)&=&\frac{3\alpha}{64\pi(G_F M_\eta^2)^2}
\left|\sum_{k=1}^3h_{ik}h_{jk}
F_2\left(\frac{M_k}{M_\eta}\right)\right|^2 \nonumber \\
&\simeq&8\times 10^{-7}\left|\sum_{k=1}^3h_{ik}h_{jk}
F_2\left(\frac{M_k}{M_\eta}\right)\right|^2,
\label{lfv}
\end{eqnarray}
where $M_\eta=1$~TeV is used and $F_2(x)$ is given by
\begin{equation}
F_2(x)=\frac{1-6x^2+3x^4+2x^6-6x^4\ln x^2}{ 6(1-x^2)^4}.
\end{equation}
Here we note that $F_2(x)\simeq \frac{1}{3x^2}$ for $x\gg 1$ and 
the present upper bounds for $Br(\mu\rightarrow e\gamma)$ 
and $Br(\tau\rightarrow \mu\gamma)$ 
are given as $5.7\times 10^{-13}$ \cite{expmeg} and $4.4\times 10^{-8}$
\cite{exptmg}, respectively.
Since $M_k> M_\eta$ is assumed in the present model, 
the bounds for these flavor violating processes give no
substantial constraint on neutrino Yukawa couplings as found 
from eqs.~(\ref{nmass}) and (\ref{lfv}). 

\subsection{Baryon number asymmetry}
Reheating process should follow the inflation discussed 
in the previous section.
In this scenario, reheating is expected to occur through the decay of $S_1$
after the inflaton stops its evolution along the above mentioned 
spiral-like trajectory and $S_{\pm 1}$ starts to oscillate 
around a global minimum of the potential. 
Although preheating could occur via scalar quartic couplings 
in the first line of eq.~(\ref{model1}), 
the reheating is expected to be finally completed 
through the decay of $S_1$ \cite{parares,reheat}. 
Since lepton number asymmetry is not produced through the particle 
creation in the preheating, we focus our study on the decay of $S_1$ here.

The decay of $S_1$ is induced by the interaction of $S_1$ 
with $\phi$ and $\eta$ during the oscillation induced by the mass terms 
which are given in the second line of eq.~(\ref{model1}).
The reheating temperature may be estimated by using the usual instantaneous 
thermalization approximation. 
If we use this approximation, the reheating temperature
is determined through the condition $H\simeq \Gamma_{\pm 1}$. 
$H$ is the Hubble parameter and $\Gamma_{\pm 1}$ stands for 
the decay width of $S_{\pm 1}$ which is the real and imaginary 
component of $S_1$.
Since $\Gamma_{\pm 1}$ can be approximately estimated as 
$\Gamma_{\pm 1}\simeq \frac{1}{8\pi}\frac{|\mu_1|^2}{m_{\pm 1}}$ where
$m_{\pm\alpha}^2=\tilde m_{S_\alpha}^2\pm m_{S_\alpha}^2$, 
the decay products of $S_{\pm 1}\rightarrow \eta\phi^\dagger,~\eta^\dagger\phi$ 
finally make thermal plasma with possible reheating 
temperature\footnote{
In this estimation, the oscillation energy of each component is assumed
to dominate the total energy density of the Universe.} \cite{reheat}
\begin{equation}
T_R^{(\pm)}\simeq 0.35g_\ast^{-1/4}
|\mu_1|\left(\frac{M_{\rm pl}}{m_{\pm 1}}\right)^{\frac{1}{2}},
\label{tr}
\end{equation}
where we use $g_\ast=116$ as the relativistic degrees of freedom 
in this model. If we consider a situation such that 
$S_{\pm \alpha}$ is not thermally generated through the inverse decay 
or the scatterings, $m_{\pm\alpha}>T_R^{(+)}$ should be satisfied at least. 
This condition could be expressed as
\begin{equation}
\frac{\mu_1}{m_{+1}}<1.9\times 10^{-4}
\left(\frac{m_{\pm\alpha}}{m_{+1}}\right)
\left(\frac{m_{+1}}{10^9~{\rm GeV}}\right)^{\frac{1}{2}}.
\label{condms}
\end{equation} 
In the following part, we confine our study to the case 
where this condition is satisfied.

The inflaton decay is relevant to 
the generation of baryon number asymmetry in this model.
The lepton number asymmetry could be directly generated through 
this process non-thermally since this decay violates the lepton number. 
In fact, if $\mu_\alpha$ is complex, the cross term 
between tree and one-loop diagrams for the decay 
could bring about the $CP$ asymmetry.
The $CP$ asymmetry induced through this decay of $S_{\pm 1}$ 
can be estimated as\footnote{In the following study, we assume 
the maximum $CP$ phase $|\sin 2(\theta_1-\theta_2)|=1$.}
\begin{eqnarray}
\epsilon_\pm&\equiv&\frac{\Gamma(S_{\pm 1}\rightarrow\eta\phi^\dagger)
-\bar\Gamma(S_{\pm 1}\rightarrow\eta^\dagger\phi)}
{\Gamma(S_{\pm 1}\rightarrow\eta\phi^\dagger)
+\bar\Gamma(S_{\pm 1}\rightarrow\eta^\dagger\phi)} \nonumber \\
&=&\pm \frac{|\mu_2|^2\sin 2(\theta_1-\theta_2)}
{16\pi}\left(\frac{1}{m_{\pm 1}^2}\ln\frac{(m_{\pm 1}^2+m_{+2}^2)m_{-2}^2}
{(m_{\pm 1}^2+m_{-2}^2)m_{+2}^2}  \right. \nonumber \\
&+& \frac{m_{\pm 1}^2-m_{+2}^2}{(m_{\pm 1}^2-m_{+2}^2)^2+m_{+2}^2\Gamma_{+2}^2}
-\left.\frac{m_{\pm 1}^2-m_{-2}^2}{(m_{\pm 1}^2-m_{-2}^2)^2
+m_{-2}^2\Gamma_{-2}^2}\right),
\label{cp}
\end{eqnarray}
where $\theta_i={\rm arg}(\mu_i)$ and 
$\Gamma_{\pm\alpha}=\frac{|\mu_\alpha|^2}{8\pi m_{\pm\alpha}}
\left(1-\frac{M_\eta^2}{m_{\pm\alpha}^2}\right)$.
As long as the condition (\ref{condms}) is satisfied, the lepton number 
asymmetry generated through the inflaton decay could be the only source
for the baryon number asymmetry since there is no mother particles 
$S_{\pm\alpha}$ in the thermal bath. 

If both components $S_{\pm 1}$ have finely degenerate masses 
$m_{+1}^2\simeq m_{-1}^2$, their decay occurs almost 
simultaneously and then $T_R^{(+)}\simeq T_R^{(-)}$.
We also find that $\epsilon_+ \simeq -\epsilon_-$ is satisfied.
Since the lepton number asymmetry generated in the $\eta$ sector 
through this decay could be estimated as $\Delta L\simeq 
\epsilon_+n_{S_{+1}}(T_R^{(+)})+\epsilon_-n_{S_{-1}}(T_R^{(-)})$,
$\Delta L$ may not take a large value in this case 
because of the cancellation
due to $\epsilon_-n_{S_{-1}}(T_R^{(-)})\simeq -\epsilon_+n_{S_{+1}}(T_R^{(+)})$.
On the other hand, if substantial mass splitting 
appears between the components $S_{\pm 1}$ and then $m_{+1}^2 > m_{-1}^2$ 
is satisfied, the $S_{+1}$ decay is expected to
occur later compared with the decay of $S_{-1}$ because of 
$\Gamma_{-1}>\Gamma_{+1}$.
In such a case, a part of lepton number asymmetry 
generated by the $S_{-1}$ decay could be washed out by the lepton number 
violating processes before the delayed $S_{+1}$ decay.
Thus, the lepton number asymmetry expected in the $\eta$ sector after
the $S_{+1}$ decay could be estimated as 
$\Delta L\simeq\epsilon_+n_{S_{+1}}(T_R^{(+)})
+{\cal K}_w(T_R^{(+)})\epsilon_-n_{S_{-1}}(T_R^{(-)})$
where ${\cal K}_w(T_R^{(+)})$ represents the washout effects from 
$T_R^{(-)}$ to $T_R^{(+)}$. 
If the lepton number violating processes decouple and then
${\cal K}_w=1$ is satisfied in this period, $\Delta L$ is expected 
to take a substantial value because 
$\epsilon_-n_{S_{-1}}(T_R^{(-)})\not= -\epsilon_+n_{S_{+1}}(T_R^{(+)})$ is satisfied.

\begin{figure}[t]
\begin{center}
\epsfxsize=7.5cm
\leavevmode
\epsfbox{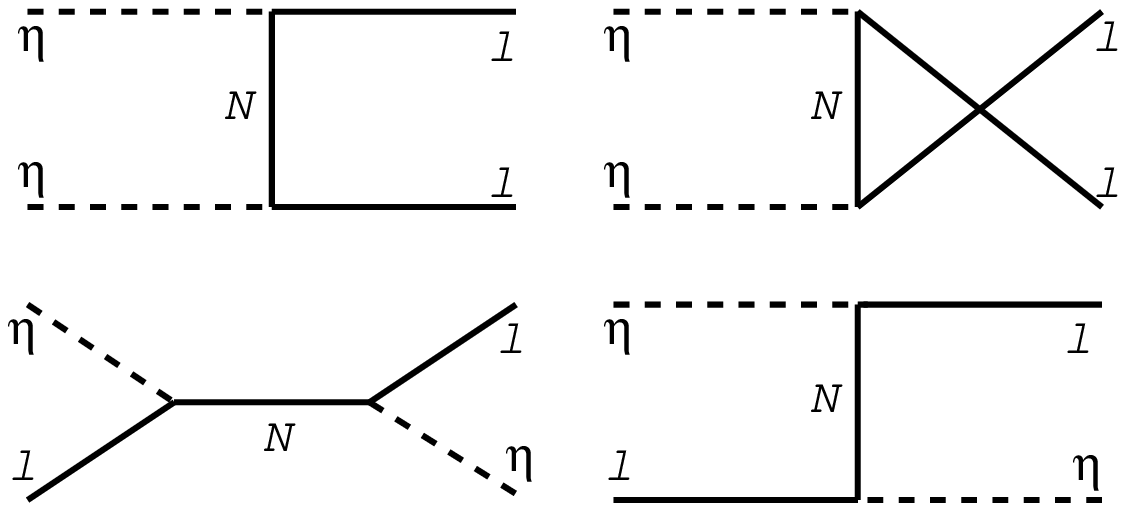}
\hspace*{5mm}
\epsfxsize=7.5cm
\leavevmode
\epsfbox{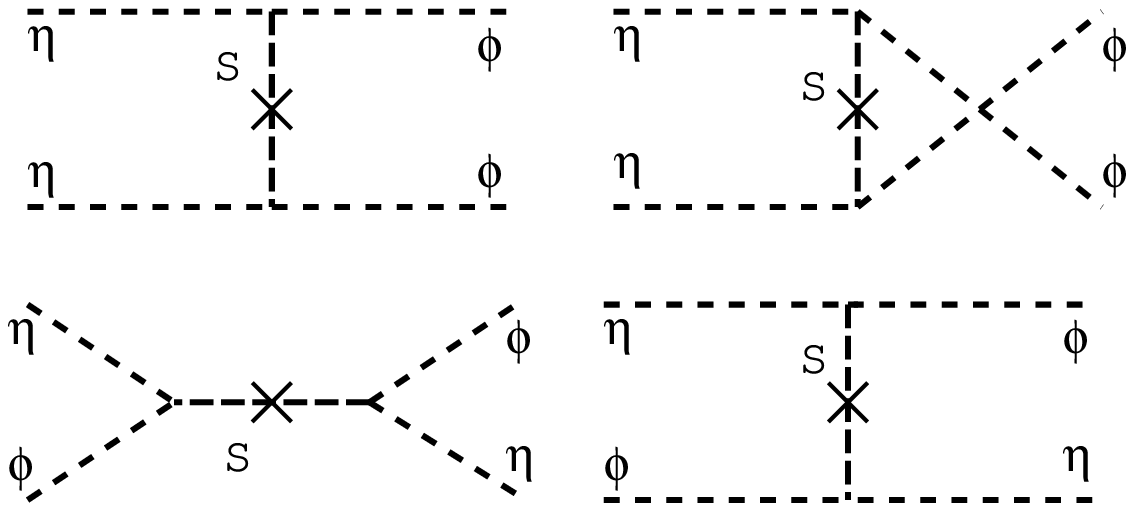}
\end{center}
\vspace*{-3mm}

{\footnotesize {\bf Fig.~2}~~Feynman diagrams which contribute to 
the transfer and the washout of the lepton number asymmetry.
The left diagrams are lepton number conserving scattering 
processes whose reaction densities are represented by $\gamma_a$
(upper ones) and $\gamma_b$ (lower one).  
The right diagrams are lepton number violating 
scattering processes whose reaction densities are represented 
by $\gamma_x$ (upper ones) and $\gamma_y$ (lower one), respectively. }
\end{figure}

The lepton number asymmetry generated in the $\eta$ sector
via the $S_{\pm 1}$ decay cannot be transferred to the SM contents through 
the decay of $\eta$. 
We should note that $\eta$ does not have any decay modes 
to the SM contents because of the $Z_2$ symmetry. However,
it could be partially transferred to the lepton sector through 
the lepton number conserving scatterings $\eta\eta\rightarrow\ell\ell$ 
and $\eta\bar\ell\rightarrow\eta^\dagger\ell$. 
These are induced by neutrino Yukawa couplings and their diagrams 
are given in the left-hand side of Fig.~2. 
On the other hand, it could also be
washed out through the lepton number violating scattering processes 
$\eta\eta\rightarrow\phi\phi$ and 
$\eta\phi^\dagger\rightarrow\eta^\dagger\phi$.
These are caused by the $S_{\pm\alpha}$ exchange due 
to the $\mu_\alpha$ couplings.
Their diagrams are also shown in the right-hand side of Fig.~2.   
In the situation where these processes are competing with each other 
before reaching the weak scale, 
the lepton number asymmetry kept in the lepton sector 
could be converted to the baryon number asymmetry through 
the sphaleron interaction.
We examine this scenario quantitatively by solving relevant Boltzmann 
equations.

For this purpose, we define the lepton number asymmetry in the co-moving 
volume as $\Delta Y_\ell\equiv\frac{n_\ell-n_{\bar\ell}}{s}$ 
in the lepton sector and 
$\Delta Y_\eta\equiv\frac{n_\eta-n_{\eta^\dagger}}{s}$ in the $\eta$ sector, 
respectively.
The entropy density $s$ is expressed as $s=\frac{2\pi^2}{45}g_\ast T^3$. 
As discussed in the previous part,
the lepton number asymmetry in the $\eta$ sector is expected 
to be fixed through the decay of $S_{\pm 1}$. 
Thus, at the reheating temperature $T_R^{(+)}$,
the lepton number asymmetry in each sector are supposed 
to be $\Delta Y_\ell(T_R^{(+)})=0$ and 
$\Delta Y_\eta(T_R^{(+)})= \frac{\epsilon_+n_{S_{+1}}(T_R^{(+)})+
\epsilon_-n_{S_{-1}}(T_R^{(-)})}{s_R}$ where $s_R$ stands for the entropy 
density at $T_R^{(+)}$.
If we use $n_{S_{\pm 1}}(T_R^{(\pm)})=\frac{\rho_{S_{\pm 1}}(T_R^{(\pm)})}{m_{\pm 1}}$ and 
$\rho_{S_{\pm 1}}(T_R^{(\pm)})=\frac{\pi^2}{30}g_\ast T_R^{(\pm )4}$
which are derived by assuming the instantaneous thermalization after
the $S_{\pm 1}$ decay, we find that the latter can be expressed as
\begin{equation}
\Delta Y_\eta(T_R^{(+)})=\frac{3}{4}\epsilon_+\frac{T_R^{(+)}}{m_{+1}}
+\frac{3}{4}\epsilon_-\frac{T_R^{(-)}}{m_{-1}}.
\label{vinit}
\end{equation} 

By taking account of the relevant processes which are explained above,
Boltzmann equations which describe the evolution of $\Delta Y_\eta$ and
$\Delta Y_\ell$ are given as\footnote{Following the usual convention, 
we introduce a dimensionless parameter $z$ as $z=\frac{M_\eta}{T}$ by using a
convenient mass scale $M_\eta$, which is defined below eq.~(\ref{nmtr2}).}
\begin{eqnarray}
\frac{d\Delta Y_\eta}{dz}&=&-\frac{z}{sH(M_\eta)}\left[
2(\gamma_a+\gamma_b)\left(\frac{\Delta Y_\eta}{Y_\eta^{\rm eq}}-
\frac{\Delta Y_\ell}{Y_\ell^{\rm eq}}\right)+
2(\gamma_x+\gamma_y)\frac{\Delta Y_\eta}{Y_\eta^{\rm eq}}\right],\nonumber \\
\frac{d\Delta Y_\ell}{dz}&=&\frac{z}{sH(M_\eta)}
2(\gamma_a+\gamma_b)\left(\frac{\Delta Y_\eta}{Y_\eta^{\rm eq}}-
\frac{\Delta Y_\ell}{Y_\ell^{\rm eq}}\right).
\label{bqn}
\end{eqnarray} 
Since we consider the case where the condition (\ref{condms}) is satisfied, 
the effect of $S_{\pm\alpha}$ in the thermal bath can be neglected.  
Each reaction density $\gamma_i$ is explained in the 
caption of Fig.~2 and their formulas are given in Appendix B. 
The generated baryon number asymmetry could be estimated as \cite{inf-nradm}
\begin{equation}
Y_B=-\frac{7}{19}\Delta Y_\ell(z_{EW})
\label{bform}
\end{equation}  
by using the lepton number asymmetry $\Delta Y_\ell$ 
obtained as the solution of these equations at the weak scale. 

Although detailed analysis of the generated baryon number asymmetry
requires to solve the above Boltzmann equations numerically, 
we briefly discuss their qualitative aspects before proceeding to it.
At first, we note the behavior of the ratio of the reaction rate 
$\Gamma$ to Hubble parameter $H$ for the relevant scattering 
processes in the case $m_{\pm\alpha}>T_R^{(+)}$
which we consider here.
$\Gamma$ and $H$ are expressed as 
$\Gamma_{a,b}\equiv\frac{\gamma_{a,b}}{n_\ell^{\rm eq}}$,
$\Gamma_{x,y}\equiv\frac{\gamma_{x,y}}{n_\eta^{\rm eq}}$ where
$n_\ell^{\rm eq}\simeq\frac{3.6M_\eta^3}{\pi^2}z^{-3}$, 
$n_\eta^{\rm eq}\simeq\frac{2M_\eta^3}{\pi^2}z^{-1}K_2(z)$
and $H(z)\simeq 0.33g_\ast^{1/2}\frac{M_\eta^2}{M_{\rm pl}}z^{-2}$.
In the lepton number conserving scattering processes 
caused by the neutrino Yukawa couplings, 
$\frac{\Gamma_a+\Gamma_b}{H}$ is a convex function 
of $z$ which takes a maximum value around $z_m\simeq \frac{M_\eta}{M_k}$.
They freeze out at $z_f(>z_m)$ in the case where 
$\frac{\Gamma_a+\Gamma_b}{H}>1$ is satisfied at $z_m$.
It is important to note that
$\Delta Y_\ell$ follows $\Delta Y_\eta$ to be $\Delta Y_\ell=\Delta Y_\eta$
as long as $\frac{\Gamma_a+\Gamma_b}{H}~{^>_\sim}~1$ is satisfied.  
On the other hand, the coupling $\mu_\alpha$ which causes 
the lepton number violating scatterings is dimensionful so that 
$\frac{\Gamma_x+\Gamma_y}{H}$ increases monotonically with $z$ throughout
the range $\frac{M_\eta}{T_R^{(+)}}<z<1$. 
Since these processes are expected to be in the thermal equilibrium 
at a certain period $z_e$ where $\frac{\Gamma_x+\Gamma_y}{H(z_e)}= 1$ 
is satisfied, $\Delta Y_\eta$ is expected to be erased at $z~{^>_\sim}~z_e$.
However, these processes are suppressed at $z~{^>_\sim}~1$ by the 
Boltzmann factor.    

Here we note that both $z_f$ and $z_e$ are determined by the
parameters relevant to the neutrino masses.
We could make a rough estimation of favored parameters for 
the generation of baryon number asymmetry
by taking account of it and the above arguments. 
As seen in eq.~(\ref{nmass}), the neutrino oscillation data imposes 
a relation for neutrino Yukawa couplings and a GeV unit $M_k$ such that 
\begin{equation}
\frac{(hh^T)_{kk}}{M_k}
\sum_{\alpha=1,2}\left(\frac{\mu_\alpha^2}{m_{-\alpha}^2}-
\frac{\mu_\alpha^2}{m_{+\alpha}^2}\right)\sim O(10^{-14}),
\label{nmassc}
\end{equation}  
where we assume $M_\eta=1$~TeV.
If we use this condition, both $z_f$ and $z_e$ can be roughly estimated as
\begin{eqnarray}
&&z_f\sim O(10^{18})\sum_k\frac{(hh^T)_{kk}^2}{M_k^2}
\sim O(10^{-11})\left[\sum_\alpha\left(\frac{\mu_\alpha^2}{m_{-\alpha}^2}-
\frac{\mu_\alpha^2}{m_{+\alpha}^2}\right)\right]^{-2}, \nonumber \\ 
&&z_e\sim O(10^{-13})\left[\sum_{\alpha=1,2}\left(\frac{\mu_\alpha^2}{m_{-\alpha}^2}-
\frac{\mu_\alpha^2}{m_{+\alpha}^2}\right)\right]^{-2},
\label{zef}
\end{eqnarray}
where the $CP$ phases of neutrino Yukawa couplings are neglected.
These results suggest that $z_f>z_e$ is always satisfied. 

The washout factor ${\cal K}_w(z)$ which we have already introduced in the
previous discussion is characterized as a decreasing 
function at $z~{^>_\sim}~z_e$ and ${\cal K}_w(z)\simeq 1$ at $z~{^<_\sim}~z_e$.
If we use it, the total lepton number at $z$ might be written as 
\begin{equation}
\Delta Y_\ell(z)+\Delta Y_\eta(z)= {\cal K}_w(z)\Delta Y_\eta(z_R),    
\label{rel1}
\end{equation}
where we use eq.~(\ref{vinit}) as the initially generated lepton number 
asymmetry.
On the other hand, the lepton number asymmetry in both sector at $z$ 
could be related as
\begin{equation}
\Delta Y_\ell(z)={\cal K}_t(z)\Delta Y_\eta(z),    
\label{rel2}
\end{equation}
where ${\cal K}_t(z)$ stands for the transfer efficiency of the
lepton number asymmetry from the $\eta$ sector to the doublet lepton sector.
If the lepton number conserving scattering processes 
are in the thermal equilibrium, ${\cal K}_t(z)= 1$ is satisfied.  
Using these relations, we could consider two possible cases 
for the generation of lepton number asymmetry in the lepton sector. 

\begin{figure}[t]
\small
\begin{center}
\begin{tabular}{c|ccccccccc}\hline
 & $h_2$ & $h_3$ & $\frac{m_{S_1}}{\tilde m_{S_1}}$ 
& $\frac{|\mu_1|}{\tilde m_{S_1}}$ 
& $\frac{\tilde m_{S_2}}{\tilde m_{S_1}}$ & $\frac{m_{S_2}}{\tilde m_{S_2}}$ & 
$\frac{|\mu_2|}{\tilde m_{S_2}}$ 
& $\epsilon_+$ & $|Y_B|$ \\ \hline 
(a)&$1.0\cdot 10^{-2}$&$4.8\cdot 10^{-3}$ & $0.5$ & $2.0\cdot 10^{-5}$  
& $1.3$ &  $0.5$  &  $3.0\cdot 10^{-3}$ 
& $1.7\cdot 10^{-5}$ & $1.0\cdot 10^{-10}$ \\  \hline 
(b)&$1.8\cdot 10^{-2}$&$9.5\cdot 10^{-3}$& $0.5$ & $ 10^{-6}$ 
& $1.3$ & $0.5$& $8.0\cdot 10^{-2}$ 
& $1.2\cdot 10^{-2}$ & $3.1\cdot 10^{-9}$  \\  \hline
\end{tabular}
\end{center}
\vspace*{3mm}

{\footnotesize {\bf Table 2.}~ The $CP$ asymmetry $\epsilon_+$ and the baryon 
number asymmetry $|Y_B|$ obtained in the present scenario for typical 
parameter settings. The dimensionful model parameters are taken to be
(a)~$M_2=2\times 10^4$, $M_3=5\times 10^4$ and $\tilde m_{S_1}=10^9$,
(b)~$M_2=2\times 10^8$, $M_3=5\times 10^8$ and $\tilde m_{S_1}=10^9$
in a GeV unit, respectively. 
Neutrino Yukawa couplings are numerically determined for $M_\eta=1$~TeV so as 
to realize the neutrino mass eigenvalues required from the neutrino 
oscillation data.}
\end{figure}

\normalsize
(a) If the lepton number conserving scatterings are in the thermal 
equilibrium at an early stage and freeze out at $z_f$,    
the lepton number asymmetry in the lepton sector at the weak scale 
is found to be roughly expressed as 
\begin{equation}
\Delta Y_\ell(z_{EW})\simeq \frac{{\cal K}_t(z_f){\cal K}_w(z_f)}
{1+{\cal K}_t(z_f)}\Delta Y_\eta(z_R).
\label{lasym}
\end{equation}
Although ${\cal K}_w(z_f)=1$ is satisfied for $z_f<z_e$, the neutrino 
mass condition allows only the situation $z_f>z_e$ as shown 
in eq.~(\ref{zef}).
Thus, the required value of $\Delta Y_\ell(z_{EW})$ could be 
obtained in the case where ${\cal K}_w(z_f)$ is not so small.
It could be realized only for $M_k \gg M_\eta$.

(b) If the lepton number conserving scattering processes never reach 
the thermal equilibrium at $z (<z_e)$ but $\frac{\Gamma_a+\Gamma_b}{H}$ has 
non-negligible values, the situation becomes completely different 
from the case (a). In this case, a part of $\Delta Y_\eta$ could 
be transferred to the lepton sector.
Since $\Delta Y_\eta$ steeply decreases at $z\sim z_e$, $\Delta Y_\ell$
could take a fixed value which might be roughly estimated as
$\Delta Y_\ell(z_e)$ independently of the value of 
$\frac{\Gamma_a+\Gamma_b}{H}$ at $z (>z_e)$.  
The transferred lepton number asymmetry $\Delta Y_\ell(z_e)$ is 
kept until the weak scale. Thus, $\Delta Y_\ell(z_{\rm EW})$ could 
be expressed as 
\begin{equation}
\Delta Y_\ell(z_{\rm EW})\simeq {\cal K}_t(z_e)\Delta Y_\eta(z_R). 
\end{equation}
where ${\cal K}_t(z_e)\ll 1$. Thus, the required lepton number 
asymmetry in the lepton sector could be obtained at the weak scale 
for a suitable ${\cal K}_t(z_e)$. 
Such a situation could happen only in the case $M_k \gg T_R^{(+)}$.

Now we present results of the numerical analysis of the Boltzmann equations.
Model parameters used in this analysis are summarized in Table 2, which 
are numerically fixed to satisfy the conditions for the neutrino masses. 
If we take account of the conditions (\ref{nmass}) and (\ref{condms}),
we find that $\frac{|\mu_1|^2}{m_{\pm 1}}\ll \frac{|\mu_2|^2}{m_{\pm 2}}$
should be satisfied and also their phases can be fixed as 
$\theta_1\not=0$ and $\theta_2=0,\frac{\pi}{2}$. 
This justifies the estimation 
in eq.~(\ref{nmass}), (\ref{nmassc}) and (\ref{zef}) and
the assumption for the maximum $CP$ phase in eq.~(\ref{cp}), which
is used in this analysis.
It also allows $\eta_R$ and $\eta_I$ to be the mass eigenstates 
of the neutral components of $\eta$. 
This becomes important for the study of DM phenomenology in the
next subsection.  

\begin{figure}[t]
\begin{center}
\epsfxsize=7cm
\leavevmode
\epsfbox{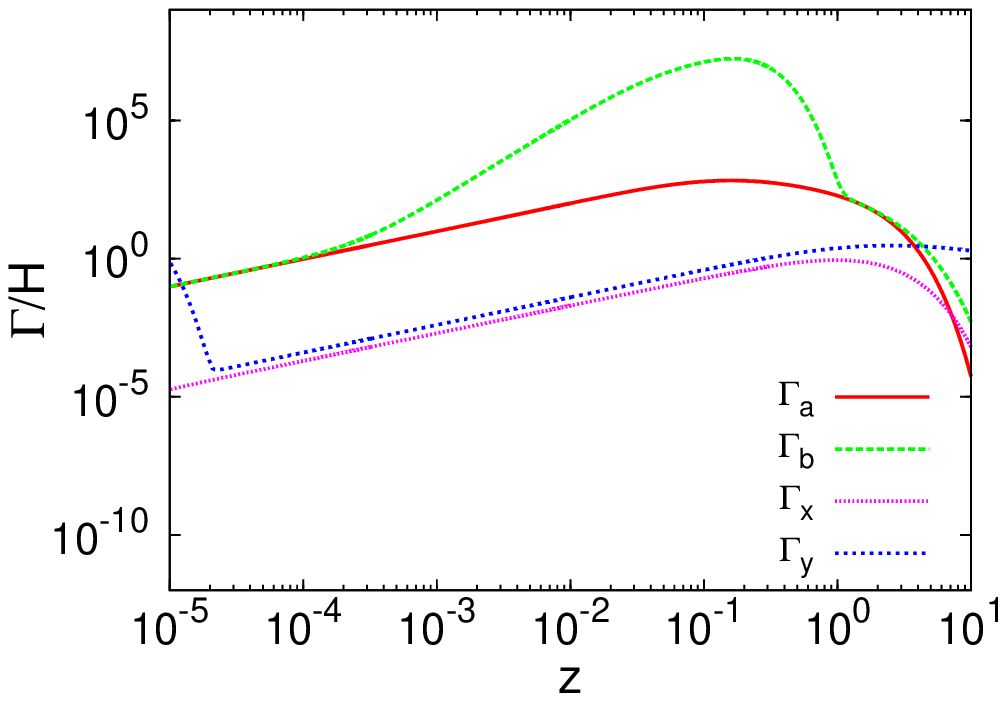}
\hspace*{5mm}
\epsfxsize=7cm
\leavevmode
\epsfbox{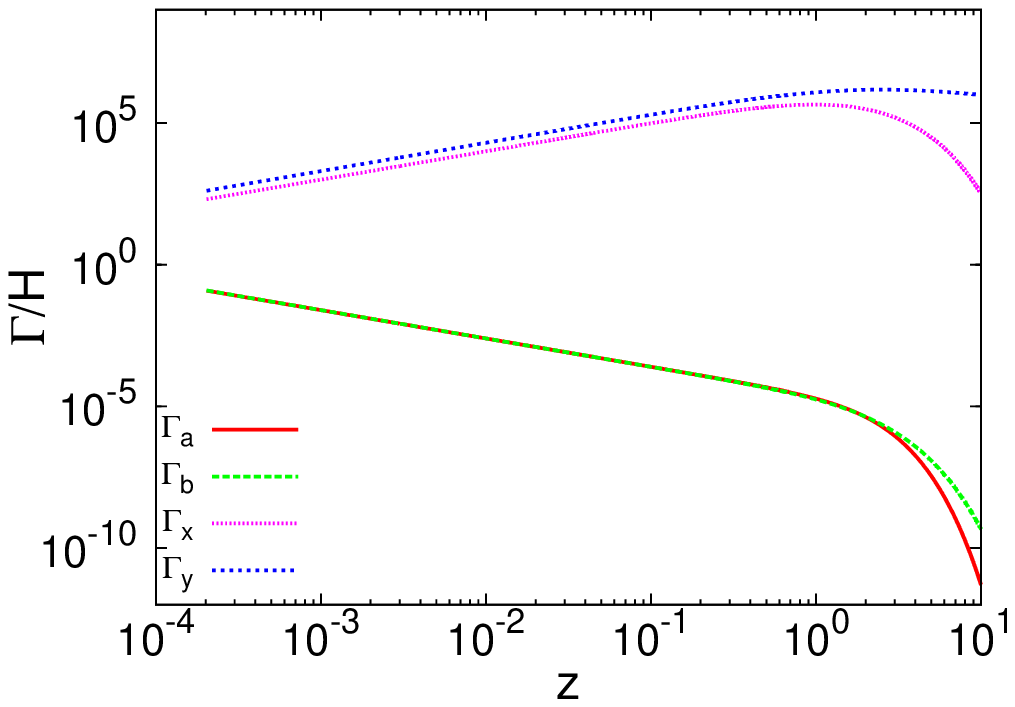} \\
\epsfxsize=7cm
\leavevmode
\epsfbox{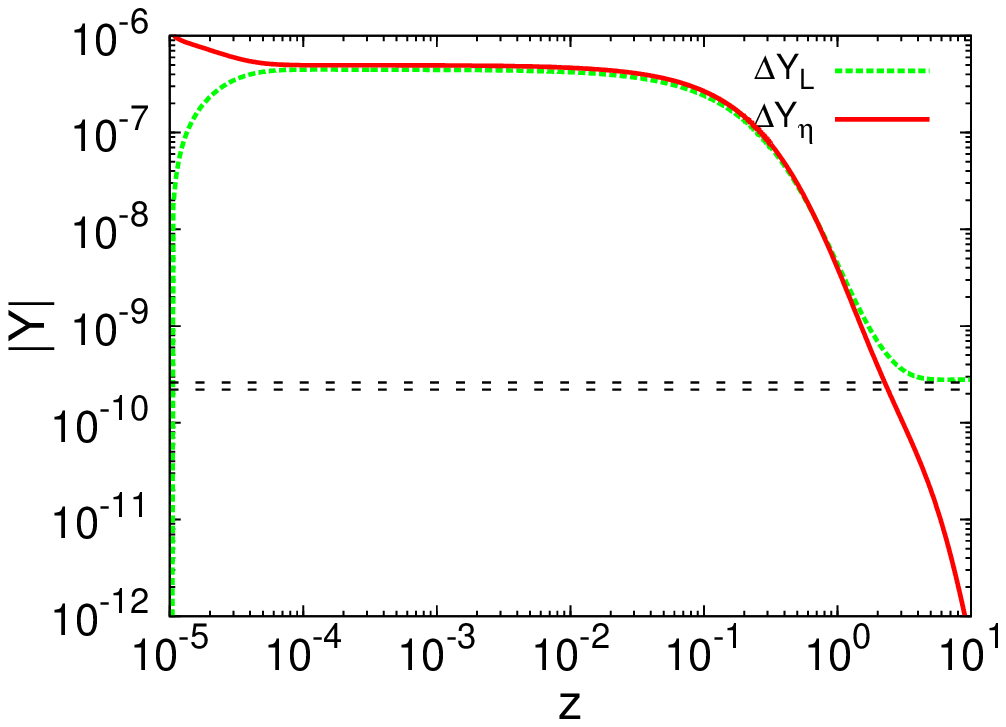}
\hspace*{5mm}
\epsfxsize=7cm
\leavevmode
\epsfbox{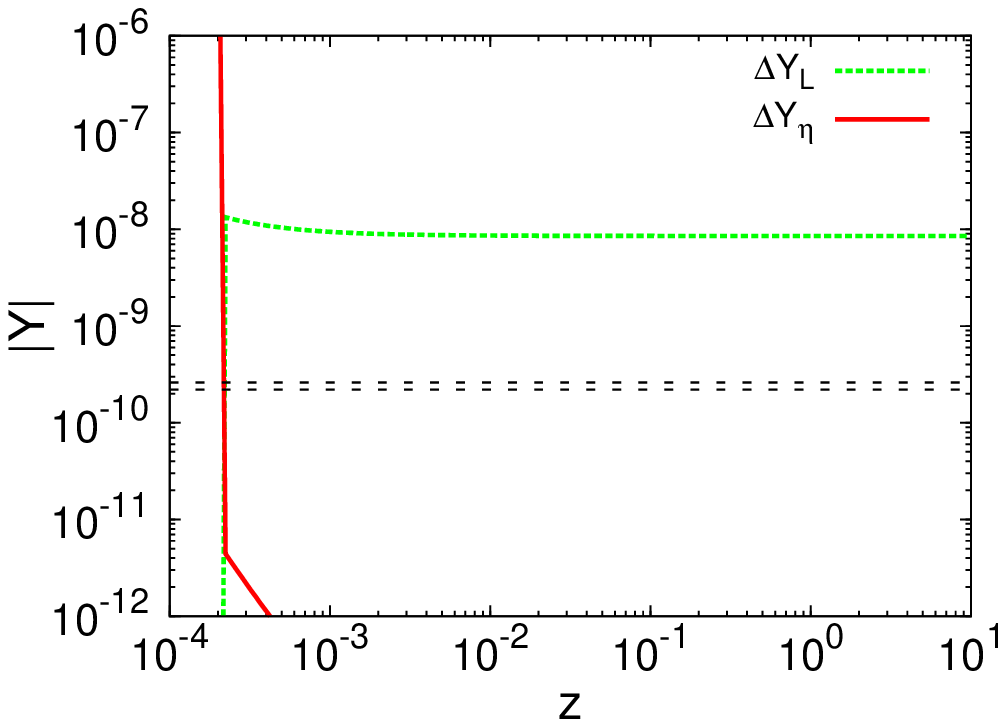}
\end{center}
\vspace*{-3mm}

{\footnotesize {\bf Fig.~3}~~The left-hand panels show the results for 
the case (a). The ratio of reaction rate $\Gamma$ to the Hubble 
parameter $H$ for each relevant process is plotted as functions of $z$ in 
the upper panel. The solutions $\Delta Y_\eta$ and $\Delta Y_\ell$ of 
the Boltzmann equations are shown as functions of $z$ in the lower panel. 
The lepton number asymmetry required to explain the observational results 
is shown by the horizontal black line.
The right-hand panels show the results for the case (b) in the same way 
as the case (a).}
\end{figure}
   
Solutions of the Boltzmann equations (\ref{bqn}) for these parameter 
settings are presented in Fig.~3.
In the upper panels of this figure, the $\frac{\Gamma}{H}$ for the 
relevant processes are plotted as functions of $z$. 
In the lower panels, $\Delta Y_\eta$ and 
$\Delta Y_\ell$ are plotted as functions of $z$. 
The lepton number asymmetry required for the suitable baryon 
number asymmetry is also shown by the horizontal black dotted lines 
in these panels.
The left and right panels show the results corresponding to 
the cases (a) and (b) discussed above, respectively.
They show that the above discussion describes qualitatively the 
features of the present scenario well.
Although our study here is done only for the limited parameter sets,
the results show that the scenario could generate the 
sufficient baryon number asymmetry for suitable model parameters 
in each case. Detailed study of this scenario for wider range 
of the model parameters will be given elsewhere.

We should recall again that the same parameters used here are 
closely related to several low energy phenomena. 
Although some of them have been discussed already, 
there is another one which has not been taken into account still now. 
We need to check the consistency with it to see whether 
the model works well or not. 
It is DM physics and this issue is the subject in the next part.

\subsection{Dark matter}
The DM candidate is built in the model as the lightest $Z_2$ odd field.
We identify it as the lightest neutral component of $\eta$.
We choose $\mu_2^2$ to be real and 
$\frac{|\mu_1|^2}{m_{\pm 1}}\ll\frac{|\mu_2|^2}{m_{\pm 2}}$ is supposed 
to be satisfied. 
In this case, the real and imaginary parts of the neutral component of $\eta$, 
which are written as $\eta_R$ and $\eta_I$, become the mass eigenstates 
as mentioned before. 
If $\eta_R$ is supposed to be a DM candidate,
$\eta_R$ could be scattered with nuclei inelastically to $\eta_I$.
It is mediated by the $Z$ boson exchange. Since it contributes to 
the DM direct search experiment \cite{inela}, 
a strong constraint is imposed on the mass difference 
$\delta(\equiv M_{\eta_I}-M_{\eta_R})$ between $\eta_R$ and $\eta_I$\footnote{The mass of $\eta_R$ and $\eta_I$ can be expressed as $M_{\eta_R}^2=M_\eta^2+\lambda_5\langle\phi\rangle^2$ and $M_{\eta_I}^2=M_\eta^2-\lambda_5\langle\phi\rangle^2$ respectively, by using the effective coupling $\lambda_5$.} \cite{ks}. 
This might give the scenario an interesting chance 
for giving a prediction in the DM direct search experiments as seen below. 

We recall the experimental situation that we have no evidence in the DM 
direct search experiments \cite{dsdm}.
If we apply it to the above mentioned process, we could put a bound 
for $\delta$. It might be estimated as $\delta> 150$~keV conservatively.
Since this mass difference is expressed in the present model as
\begin{equation}
\delta\simeq \frac{\langle\phi\rangle^2}{M_\eta}
\left(\frac{\mu_2^2}{m_{-2}^2}-\frac{\mu_2^2}{m_{+2}^2}\right),
\end{equation} 
the constraint is found to be represented as
\begin{equation}
\left(\frac{\mu_2^2}{m_{-2}^2}-\frac{\mu_2^2}{m_{+2}^2}\right)
~{^>_\sim}~5\times 10^{-6}\left(\frac{M_\eta}{1~{\rm TeV}}\right).
\label{inelas}
\end{equation}
As noted in the previous part, 
the left-hand side of eq.~(\ref{inelas}) corresponds to the effective 
coupling $|\lambda_5|$ for the assumed parameters. 
Although this constraint depends on the DM velocity 
distribution in our galaxy and other uncertain factors, eq.~(\ref{inelas})
gives an interesting condition for the present scenario
on the origin of the baryon number asymmetry.
We find that the model parameters used in the case (b) gives 
$|\lambda_5|\sim 3\times 10^{-3}$ and then this condition is clearly satisfied. 
On the other hand, the situation is subtle in the case (a) 
since we find $|\lambda_5|\sim 5\times 10^{-6}$. 
This suggests that the DM candidate in this model could be detected 
through the inelastic scattering in the direct search experiments 
if this leptogenesis scenario is realized in Nature for this parameter range.
It may be worthy to reexamine the direct search results in this mass range
in detail.
  
The above scenario should be also consistent with the DM relic abundance. 
In the present study, DM is assumed to be $\eta_R$. 
In general, its relics could come from two types of origin such as
\begin{equation}
\Omega h^2=\Omega_{\rm th} h^2 +\Omega_{\rm nonth} h^2.
\end{equation} 
The first one is the usual thermal relic, that is, the remnant of $\eta_R$ 
decoupled from the thermal equilibrium distribution. 
It can be estimated by using the usual formulas \cite{relic},
\begin{equation}
\Omega_{\rm th}h^2=\frac{1.07\times 10^9z_{DM}}{g_\ast^{1/2}m_{\rm pl}({\rm
 GeV})\langle\sigma_\eta v\rangle}, \qquad
z_{DM}=\ln\frac{0.038g m_{\rm pl}M_{\eta_R}\langle\sigma_\eta v\rangle}
{g_\ast^{1/2}z_{DM}^{1/2}},
\end{equation}
where $m_{\rm pl}=\sqrt {8 \pi}M_{\rm pl}$ and $g$ is internal degrees 
of freedom of DM. $z_{DM}$ is defined 
by $z_{DM}=\frac{M_{\eta_R}}{T_f}$ for the $\eta_R$
freeze-out temperature $T_f$. The relevant thermally averaged annihilation 
cross section $\langle \sigma_\eta v\rangle$ including the 
co-annihilation processes can be found in \cite{ham,ks}. 
Since $\langle \sigma_\eta v\rangle$ has a crucial dependence on 
the couplings $\lambda_{3,4}$ given in eq.~(\ref{model}) \cite{ks}, 
the relic abundance $\Omega_{\rm th}h^2$ could change its value 
by varying the values of $\lambda_{3,4}$ without 
affecting other phenomena discussed in this paper. 
Thus, it is not difficult to realize the suitable relic abundance 
from this source.   

The second one comes from the non-thermal origin, that is,
the lepton number asymmetry left in the $\eta$ sector 
which is produced through the decay of $S_{\pm 1}$.
One may consider that this could play an important role for 
the DM relic abundance as in the asymmetric DM scenario. 
In fact, its contribution could be estimated as
\begin{equation}
\Omega_{\rm nonth}h^2=2.8\times 10^{11}
\left(\frac{M_\eta}{1~{\rm TeV}}\right)\Delta Y_\eta,
\end{equation}
where $\Delta Y_\eta$ is the asymmetry in the present Universe.
The non-negligible contribution to the DM relic abundance 
is expected in the case $\Delta Y_\eta=O(10^{-13})$.
However, we should note that the relic abundance of 
$\eta_R$ is fixed after the electroweak symmetry breaking. 
Since the lepton number in the $\eta$ sector is violated 
through the $\eta_R$-$\eta_I$ mass splitting caused by the 
electroweak symmetry breaking mediated by the effective 
coupling $\lambda_5$, the lepton number asymmetry in the $\eta$ sector 
disappears completely at this stage. 
Thus, this non-thermal component cannot contribute to the 
DM relic abundance in this scenario.
The DM relic abundance is completely determined only by the thermal 
relics as in the same way discussed in the previous studies \cite{ks}.
This suggests that the leptogenesis scenario presented here 
can generate sufficient baryon number asymmetry in a consistent 
way with the generation of the neutrino masses, the DM phenomenology 
and others. 
It is notable that they are closely related to each other through 
the inflaton interaction with the SM Higgs scalar and $\eta$.
  
\section{Summary}
We have considered an extension of the radiative neutrino mass model 
with singlet scalars, one of which plays a role of inflaton. 
The original Ma model can be obtained effectively 
at low energy regions by integrating out the singlet scalars.
In this model, the lepton number violation is prepared as the mass 
term of inflaton and it plays a crucial role in both the radiative 
neutrino mass generation and the generation of the lepton number 
asymmetry.
The lepton number asymmetry is produced by the inflaton decay 
firstly in the inert doublet sector. 
It is transferred from the inert doublet sector to 
the lepton sector through the lepton number conserving scatterings. 
We have examined this scenario numerically and showed that the 
sufficient baryon number asymmetry could be generated as long as
the model parameters take suitable values.
They can be consistent with the neutrino mass generation and the 
DM phenomenology. 
The scenario could present a new possibility for the leptogenesis 
in the framework which makes a close connection between the neutrino 
mass generation and the inflation of the Universe.  

\section*{Acknowledgement}
S.~K. is supported by Grant-in-Aid for JSPS fellows (26$\cdot$5862).
D.~S. is supported by JSPS Grant-in-Aid for Scientific
Research (C) (Grant Number 24540263) and MEXT Grant-in-Aid 
for Scientific Research on Innovative Areas (Grant Number 26104009).

\newpage
\section*{Appendix A}
In this Appendix, we fix the concrete form of the neutrino mass 
matrix to determine the model parameters based on the neutrino 
oscillation data.
Since it determines the flavor structure of neutrino Yukawa
couplings, we can fix the reaction density contained in the Boltzmann 
equations. 
As such a typical example, in the present analysis we use  
\begin{equation}
h_{ei}=0,~ h_{\mu i}=h_{\tau i}\equiv h_i\quad (i=1,2)  ; \qquad 
h_{e3}=h_{\mu 3}=-h_{\tau 3}\equiv h_3 ,
\end{equation}
which could realize the tri-bimaximal neutrino mixing \cite{raddm2}. 
Although it is not realistic, it could give a good starting point          
for the purpose of this paper.
In this case, three neutrino mass eigenvalues are given as
\begin{equation}
m_{\nu_1}=0, \quad, m_{\nu_2}=3h_3^2\Lambda_3, \quad 
m_{\nu_3}=2(h_1^2\Lambda_1+h_2^2\Lambda_2),  
\end{equation}
where $\Lambda_k$ is defined by
\begin{equation}
\Lambda_k=\sum_{\alpha=1,2}\sum_{f=\pm}
\frac{M_k\mu_\alpha^{(f)2}\langle\phi\rangle^2}{8\pi^2}
I(M_\eta,M_k,m_{f\alpha}).
\end{equation} 
Thus, $m_{\nu_3}=\sqrt{\Delta m^2_{\rm atm}}$ and 
$m_{\nu_2}= \sqrt{\Delta m^2_{\rm sol}}$
should be satisfied for the normal hierarchy case. We use this relation
to fix the values of neutrino Yukawa couplings in the present analysis.

\section*{Appendix B}
In this Appendix, we give the formulas of the reaction density
contributing to the Boltzmann equations for the lepton number asymmetry.
In order to give the expression for the reaction density of the relevant
processes, we introduce dimensionless variables as
\begin{equation}
x=\frac{s}{M_\eta^2}, \qquad a_j=\frac{M_j^2}{M_\eta^2}, \qquad 
b_{\pm \alpha}=\frac{m_{\pm \alpha}^2}{M_\eta^2}, \qquad 
b_{\mu_\alpha}=\frac{|\mu_\alpha|^2}{M_\eta^2},
\end{equation}
where $s$ is the squared center of mass energy.

The reaction density for the scattering process is expressed as
\begin{equation}
\gamma(ab\rightarrow ij)=\frac{T}{64\pi^4}\int^\infty_{s_{\rm min}}ds~
\hat\sigma(s)\sqrt{s}K_1\left(\frac{\sqrt{s}}{T}\right),
\end{equation}
where $\hat\sigma(s)$ is the reduced cross section and
$K_1(z)$ is the modified Bessel function of the second kind.
The lower bound of integration is defined as 
$s_{\rm min}={\rm max}[(m_a+m_b)^2,(m_i+m_j)^2]$.  

The lepton number conserving scattering processes are induced 
by the diagrams with $N_i$ exchange which are shown in the left-hand 
side of Fig.~2.
In order to give the expression for the reaction density of these
processes, we define the following quantities for convenience:
\begin{equation}
\frac{1}{D_i(x)}=\frac{x-a_i}{(x-a_i)^2+a_i^2c_i}, \qquad 
c_i=\frac{1}{64\pi^2}\left(\sum_{k=e,\mu,\tau}
|h_{ki}|^2\right)^2\left(1-\frac{1}{a_i}\right)^4.
\end{equation}
Using these definitions, their reduced cross sections are expressed as 
\begin{eqnarray}
\hat\sigma_a(x)&=&\frac{1}{2\pi}
\left[\sum_{i=1}^3(hh^\dagger)^2_{ii}\left\{
\frac{a_i(x^2-4x)^{1/2}}{a_ix+(a_i-1)^2}\right.\right. \nonumber \\
&+&\left.\left.
\frac{a_i}{x+2a_i-2}
\ln\left(\frac{x+(x^2-4x)^{1/2}+2a_i-2}
{x-(x^2-4x)^{1/2}+2a_i-2}\right)\right\}
\right.\nonumber\\
&+&\left.
\sum_{i>j}
\frac{{\rm Re}[(hh^\dagger)_{ij}^2]\sqrt{a_ia_j}}{x+a_i+a_j-2}
\left\{
\frac{2x+3a_i+a_j-4}{a_j-a_i}
\ln\left(\frac{x+(x^2-4x)^{1/2}+2a_i-2}
{x-(x^2-4x)^{1/2}+2a_i-2}\right)\right.\right. \nonumber \\
&+&\left.\left. \frac{2x+a_i+3a_j-4}{a_i-a_j}
\ln\left(\frac{x+(x^2-4x)^{1/2}+2a_j-2}
{x-(x^2-4x)^{1/2}+2a_j-2}\right)
\right\}\right] 
\label{lv2}
\end{eqnarray}
for $\eta\eta \rightarrow \ell_\alpha\ell_\beta $ and 
\begin{eqnarray}
\hat\sigma_b(x)&=&\frac{1}{2\pi}\frac{(x-1)^2}{x^2}
\left[\sum_{i=1}^3(hh^\dagger)_{ii}^2\frac{a_i}{x}
\left\{\frac{x^2}{xa_i -1}+\frac{x}{D_i(x)}
+\frac{(x-1)^2}{2D_i(x)^2}\right.\right.\nonumber \\
&-&\left.\frac{x^2}{(x-1)^2}
\left(1+\frac{x+a_i-2}{D_i(x)}\right)
\ln\left(\frac{x(x+a_i-2)}{xa_i-1}\right)\right\}\nonumber \\
&+&\left.
\sum_{i>j}{\rm Re}[(hh^\dagger)_{ij}^2]\frac{\sqrt{a_ia_j}}{x}\left\{
\frac{x}{D_i(x)}+\frac{x}{D_j(x)}+\frac{(x-1)^2}{D_i(x)D_j(x)}
\right.\right.\nonumber \\
&+&\left.\left.\frac{x^2}{(x-1)^2}
\left(\frac{2(x+a_i-2)}{a_j-a_i}-
\frac{x+a_i-2}{D_j(x)}\right)\ln\frac{x(x+a_i-2)}{xa_i-1}
\right.\right. \nonumber\\
&+&\left.\left.\frac{x^2}{(x-1)^2}
\left(\frac{2(x+a_j-2)}{a_i-a_j}-
\frac{x+a_j-2}{D_i(x)}\right)\ln\frac{x(x+a_j-2)}{xa_j-1}
\right\}\right]
\label{lv1}
\end{eqnarray}
for $\ell_\alpha\eta^\dagger \rightarrow \bar\ell_\beta\eta$.

The lepton number violating scattering processes are brought about 
by the diagrams with $S_{\pm\alpha}$ exchange which are shown in 
the right-hand side of Fig.~2.
In order to represent their reduced cross section, 
we introduce the definition such as
\begin{eqnarray}
&&\frac{1}{\tilde D_{\pm\alpha}(x)}=\frac{1}
{(x-b_{\pm\alpha})^2+b_{\pm\alpha}^2\tilde c_{\pm\alpha}}, \qquad
\tilde c_{\pm\alpha}=\frac{1}{64\pi^2}
\left(\frac{b_{\mu_{\pm\alpha}}}{b_{\pm\alpha}}\right)^2
\left(1-\frac{1}{b_{\pm\alpha}}\right), \nonumber \\ 
&&P_{\pm \alpha}=\frac{2(1-b_{\pm \alpha})-x}{[x(x-4)]^{1/2}}, \qquad
Q_{\pm \alpha}=-1 +\frac{2(1-xb_{\pm \alpha})}{(x-1)^2}.
\end{eqnarray}
Using these quantities, the reduced cross sections are represented as 
\begin{eqnarray}
\hat\sigma_x(x)&=&\sum_{\alpha=1,2}\frac{b_{\mu_\alpha}^2}{4\pi}
\frac{1}{(x^3(x-4))^{1/2}}
\left[\frac{2}{P_{+\alpha}^2-1}+\frac{2}{P_{-\alpha}^2-1} \right. \nonumber \\
&+&\left(\frac{1}{P_{+\alpha}}+\frac{4P_{-\alpha}}{P_{+\alpha}^2-P_{-\alpha}^2}\right)
\ln\frac{P_{+\alpha}+1}{P_{+\alpha}-1} \nonumber \\
&+&\left.\left(\frac{1}{P_{-\alpha}}-
\frac{4P_{+\alpha}}{P_{+\alpha}^2-P_{-\alpha}^2}\right)
\ln\frac{P_{-\alpha}+1}{P_{-\alpha}-1}\right] \nonumber \\
&+& {\rm (cross~terms~between~\alpha=1~and~2)}
\end{eqnarray}
for $\eta\eta\rightarrow\phi\phi$ and
\begin{eqnarray}
\hat\sigma_y(x)&=&\sum_{\alpha=1,2}\frac{b_{\mu_\alpha}^2}{2\pi}\left[
\frac{1}{(x-1)^2}\left\{\frac{1}{Q_{+\alpha}^2-1}+\frac{1}{Q_{-\alpha}^2-1} 
\right.\right. \nonumber \\
&+&\left.\frac{1}{Q_{+\alpha}-Q_{-\alpha}}\left(
\ln\frac{Q_{+\alpha}+1}{Q_{+\alpha} -1}
-\ln\frac{Q_{-\alpha}+1}{Q_{-\alpha} -1}
\right)\right\} \nonumber \\
&+&\frac{(x-1)^2}{4x^2}\left\{\frac{1}{\tilde D_{+\alpha}(x)} 
+ \frac{1}{\tilde D_{-\alpha}(x)}
-\frac{2}{b_{+\alpha}-b_{-\alpha}}\left(
\frac{x-b_{+\alpha}}{\tilde D_{+\alpha}(x)}-
\frac{x-b_{-\alpha}}{\tilde D_{-\alpha}(x)}\right)\right\} 
\nonumber \\
&+&\left.\frac{1}{2x}\left( \frac{x-b_{+\alpha}}{\tilde D_{+\alpha}(x)}
-\frac{x-b_{-\alpha}}{\tilde D_{-\alpha}(x)}\right)
\left(\ln\frac{Q_{+\alpha}+1}{Q_{+\alpha} -1}
-\ln\frac{Q_{-\alpha}+1}{Q_{-\alpha} -1} \right)\right] \nonumber \\
&+& {\rm (cross~terms~between~\alpha=1~and~2)}
\end{eqnarray}
for $\eta\phi^\dagger\rightarrow \eta^\dagger\phi$.
Since we consider the case $b_{\mu_2}\gg b_{\mu_1}$, we can neglect 
contributions relevant to $b_{\mu_1}$.

\newpage
\bibliographystyle{unsrt}

\end{document}